\documentclass[12pt]{article}

\makeatletter
\@addtoreset{equation}{section}
\renewcommand{\theequation}{\thesection.\arabic{equation}}
\renewcommand{\@eqnnum}{\rm{(\theequation)}}
\makeatother

\newcommand{\claim}[1]{\medskip\noindent{\bf #1}}

\newcommand{\qq}{\begin{equation}}
\newcommand{\qqq}{\end{equation}}
\newcommand{\myeqnarray}[1]{
  \begingroup
  \jot=#1pt
  \arraycolsep=2pt
  \begin{eqnarray}}
\newcommand{\beqnarray}{\myeqnarray{3}}
\newcommand{\eeqnarray}{\end{eqnarray}\endgroup}
\newcommand{\ba}{\begin{array}{cc}}
\newcommand{\ea}{\end{array}}

\def\tilde{\widetilde}

\newcommand{\da}{\partial}

\newcommand{\un}{\underline}

\newcommand{\la}{\lambda}

\newcommand{\hf}{{_1\over^2}}

\newcommand{\vs}{\vskip}

\newcommand{\Na}{{\bf a}}

\newcommand{\NE}{{\bf E}}
\renewcommand{\NG}{{\bf G}}
\newcommand{\Ng}{{\bf g}}
\newcommand{\NU}{{\bf U}}

\newcommand{\NX}{{\bf X}}

\newcommand{\NY}{{\bf Y}}

\newcommand{\Nv}{{\bf v}}
\newcommand{\Nw}{{\bf w}}

\newcommand{\Nb}{{\bf b}}

\newcommand{\ND}{{{\bf D}}}
\newcommand{\NR}{{{\bf R}}}

\newcommand{\NP}{{{\bf P}}}

\newcommand{\NZ}{{{\bf Z}}}
\newcommand{\NH}{{{\bf H}}}
\newcommand{\NN}{{{\bf N}}}

\newcommand{\NM}{{{\bf M}}}

\newcommand{\NW}{{{\bf W}}}
\newcommand{\NV}{{{\bf V}}}

\newcommand{\NJ}{{{\bf J}}}

\newcommand{\CA}{{\cal A}}

\newcommand{\CC}{{\cal C}}

\newcommand{\CN}{{\cal N}}
\newcommand{\CO}{{\cal O}}
\newcommand{\CP}{{\cal P}}

\newcommand{\CU}{{\cal U}}

\newcommand{\CX}{{\cal X}}

\newcommand{\CZ}{{\cal Z}}

\begin{document}
\vskip 4.0cm
\begin{center}
{\large{\bf{Perturbative analysis of anharmonic chains of oscillators out of equilibrium}}}
\vskip 1.5cm

R.~Lefevere\footnote{Supported by the Japan Society for the Promotion of
Science and by Les services f\'ed\'eraux des affaires scientifiques, techniques et culturelles du gouvernement Belge .} \\ Institut de Physique Th\'eorique\\ Universit\'e Catholique de Louvain \\B-1348 Louvain-la-Neuve\\ Belgium
\vskip 1.0cm

A.~Schenkel\footnote{Supported by the Swiss National Science Foundation.}
\\Department of Mathematics\\Helsinki University\\ P.O. Box 4, 00014 Helsinki \\ Finland

\end{center}

\date{ }

\vskip 1.3 cm

\begin{abstract}
\vskip 0.3cm
We compute the first-order correction to the correlation functions of the stationary state of a
stochastically forced harmonic chain out of equilibrium when a small
on-site anharmonic potential is added. 
This is achieved
by deriving a suitable formula for the covariance matrix of the
invariant state. 
We find that the first-order correction of the heat current
does not depend on the size of the system. Second,
the temperature profile is linear when
the harmonic part of the on-site potential is zero. 
The sign of the gradient of the profile, however, is opposite to the
sign of the temperature difference
of the two heat baths.

\noindent
\end{abstract}

\vs  1.6cm
\
\section{Introduction}
The goal of this paper is to begin a perturbative analysis of invariant
probability measures arising in the context of non-equilibrium statistical mechanics.
As a model at hand, we will consider a Hamiltonian chain of $N$ oscillators interacting through
nearest-neighbour interactions, coupled at its boundaries to stochastic heat baths of different
temperatures,
and that we will perturb by a small anharmonic (quartic)
on-site interaction. The covariance of the stationary state in the purely harmonic case has been
computed in \cite{rll,nak}.  For other cases, i.e. anharmonic cases, almost nothing is
known about the physical content of the stationary state, except results about the
positivity of entropy production and validity of linear response
theory \cite{eck2}.  

It is a natural idea to attempt to understand its physical properties
by performing a perturbative
analysis. Such an approach, based on the phonon picture, has been exploited by
physicists to tackle the Fourier law, see \cite{Mermin} for a classical exposition. In
particular, the Peierls theory seems successful in computing  the thermal conductivity
and its thermal and dimensional dependence.  The Peierls approach assumes from
the beginning the existence of an infinite non-equilibrium state where
local temperature equilibrium is expected to hold.
It is also based on several implicit
assumptions, such as the validity of a Boltzmann equation for phonons.  
In this paper, we adopt a different approach and begin a rigorous perturbative analysis
of a finite (although taking $N$ large will have some simplifying
features) anharmonic chain.  
Our starting point is a formula, which seems to be new, for the correlation functions of the
stationary measure. This formula allows us to derive (matrix) equations for the first-order
correction.
The relationship between our approach by stationary non-equilibrium
states (SNS) and the Peierls approach is, 
at this stage, far from clear.
A first interesting step would be to achieve some understanding 
of the equivalence of the definition of
the thermal conductivity by the Green-Kubo formula and its definition
in the SNS approach as, roughly speaking, 
the ratio of the heat current
and the temperature gradient.

The main obstacle to developing a
perturbative expansion of SNS's
is that, in contrast to the equilibrium case, no explicit
formula for the invariant density is known.  Moreover, the fact that the relevant models
are degenerate in a stochastic sense makes it laborious to obtain a systematic perturbative expansion
starting from the equations of motion.
We circumvent this difficulty by deriving a formula for the two-point
correlation functions of invariant states,
which holds
under the assumption of $L^1$-convergence of the finite-time
correlation functions to those
of the (unique) invariant measure.
We emphasize that the validity of the formula is not restricted to
the concrete problem of the anharmonic chain considered here. 
It may prove useful whenever
the invariant measure is not explicitly known,
in particular in the context of transport phenomena modeled by hypoelliptic
stochastic processes.  We also remark
that the form of the formula for the covariance is very similar to,
and provides a lower bound on, the expectation of the Malliavin matrix.  

Our main result concerning the heat current
is that its first-order correction
remains uniformly bounded as the number of oscillators goes to
infinity. 
In particular, perturbative analysis does not, at first order, reveal
any sign that Fourier law holds in such anharmonic models as
numerical studies suggest, see e.g. \cite{huli}.  
Furthermore, we find that the first-order correction to the
temperature profile is exponentially decaying in the bulk of the
chain, with a decay rate that depends on the strength of the harmonic
part of the on-site potential.  When this strength vanishes, the
correction to the temperature profile is linear.  However, the sign is
``wrong", in the sense that the linear profile has the {\sl lowest}
temperature near the {\sl hottest} bath and the {\sl highest}
temperature near the {\sl coldest} bath.  This is analogous to the
result of \cite{rll}, where the temperature profile is also oriented
in the ``wrong" direction.  The main difference is of course that in
\cite{rll}, the temperature profile is exponentially decaying.  
In order to understand what is responsible for this awkward behaviour,
it would be interesting to examine the perturbation theory of harmonic
chains that are defined differently near their ends, e.g., with
respect to the harmonic interaction or the coupling with heat baths. 
Another feature of our solution is that the temperature profile is
shifted downwards, in the sense that the temperature at the middle
point of the chain is lower than the arithmetic mean of the
temperatures of the heat baths.

The remainder of this paper is organized as follows.  In Section 2, we
specify the basic set-up for the type of anharmonic chains we will
consider.  Section 3 is devoted to the derivation of our basic formula
for the covariance.  In Section 4, we derive the matrix equations for
the first-order corrections to the harmonic case. One assumption
of this section is that the invariant measure is regular in the
anharmonic parameter.  We postpone the proof of this fact to a future
publication.
The last two sections are devoted to the resolution of these equations.
This is done by generalizing the methods of \cite{nak,rll}.

\section{A model for heat conduction}

In order to explain the behaviour of the thermal conductivity in crystalline solids,
one often models the solid by a chain (or lattice in higher dimension) whose ends
are coupled to heat baths maintained at different temperatures.  The coupling can be
taken stochastic and more precisely of Langevin type. In one dimension, the set-up is
as follows. At each site of a lattice $\{1,\ldots,N\}$ is attached a particle
of momentum $p_i$ and position $q_i$. The dynamics is Hamiltonian in the bulk and
stochastic through the Langevin coupling to heat baths at the boundaries. The
Hamiltonian is of the form,
\qq
H(\un p,\un q)=\sum_{i=1}^{N}\Bigl(\hf p_i^2 +V(q_i)\Bigr)+\sum_{i=2}^N
U(q_i-q_{i-1})+U(q_1)+U(q_N).
\label{Ham}
\qqq
Specific choices for the potentials $U$ and $V$ will be specified below.  
The equations
of motions are given by,
\beqnarray
dq_i&=&p_idt,\quad i=1,\dots,N,
\label{eqq}
\\
dp_i&=&-\frac{\da H}{\da q_i}(\un p,\un q)dt,\quad i=2, \ldots,N-1,
\label{eqp1}
\eeqnarray
and,
\beqnarray
dp_{1}&=&-\frac{\da H}{\da q_{1}}(\un p,\un q)dt-\gamma p_{1} dt
+\sqrt{2\gamma kT_1}\,dw_{l}\,,
\label{eqp2}
\\
dp_{N}&=&-\frac{\da H}{\da q_N}(\un p,\un q)dt-\gamma p_{N} dt
+\sqrt{2\gamma kT_N}\,dw_{r}\,.
\label{eqp3}
\eeqnarray
$T_1$  and $T_N$ stand for the temperature of the left and
right reservoirs, respectively,
whereas  $w_{l}$ and $w_r$ are two independent standard Wiener processes.

It is an easy fact to check that when $T_1=T_N=T=\beta^{-1}$,
the measure on the configuration space ${\bf R}^{2N}$ whose density
with respect to the Lebesgue measure is given by
\qq
\rho(\un p,\un q)=Z^{-1}e^{-\beta H(\un p,\un q)}
\qqq
is invariant (stationary) for the stochastic dynamics defined above.
Explicitly, one can check that for $L$ the generator of the dynamics
and any function $f$ in its domain,
\qq
\int  Lf\, \rho (\un p,\un q)\,d\un p\,\un dq=0.
\label{sta}
\qqq
In the case of two different temperatures, existence, uniqueness and
exponential convergence to an unique invariant state has been established under fairly
general conditions on the potentials $U$ and $V$ \cite{eck1, eck2,eck3, rbt}. In the
case of harmonic coupling, the covariance of the stationary state has been exactly
computed in \cite{rll,nak}.

An essential ingredient
of the proof of the uniqueness is the fact that the system satisfies
the so-called H\"ormander condition.
This condition implies that the noise spreads in a
sufficiently good way through the system, so that the
transition probabilities have smooth densities.  This property is
encapsulated in the non-degeneracy of the Malliavin matrix
associated to the stochastic system under study. As the noise represents the injection of energy
into the system, it is
natural to enquire about the relationship between the Malliavin
matrix and the correlation functions of the stationary state.  This might provide a way to tackle the description of the
stationary state when its density is not explicitly known.
Indeed, from a physical point of view, the central
question, once uniqueness has been established, is to compute the
energy spectrum and correlation functions of the stationary state and ultimately, to establish the validity of the Fourier law.
As mentioned above, the case of a harmonic chain has
been completely and explicitly solved.
The main feature of the solution is
a flat temperature profile and an associated
infinite thermal conductivity.

The basic idea in order to perform a perturbation theory of the
non-equilibrium stationary state is to write the two-point correlation function of the
stationary measure under a ``Malliavin'' form, 
similar to the form derived by Nakazawa in the Gaussian
harmonic case, \cite{nak}.

\section{The Malliavin matrix and the covariance matrix of the stationary measure}
We consider now a general system of stochastic equations.
Denote by $x_t\in{\bf R}^d$ the solution of the stochastic
differential equation,
\qq
dx_t=X_0(x_t)\,dt+\sum_{k=1}^n X_k(x_t)\, dw_k(t)
\label{seq}
\qqq
with initial condition $x_0=x$, where the $w_k$'s are $n$ independent 
one-dimensional Brownian
motions and $X_l$, $l=0,\dots,n$, are $\CC^\infty$ vector fields over
${\bf R}^d$ 
satisfying for any multi-index $\alpha$,
\qq
||\partial^\alpha X_l(x)||\leq C(1+||x||^{K_\alpha})
\label{xbo}
\qqq
for some $K_\alpha>0$.  
We note that solutions to such equations are in general not ensured to exist
globally. In the sequel, we restrict ourselves to the following
situations.

\claim{Assumption 3.1.} {\sl For all $x\in\NR^d$, equation (\ref{seq})
has a unique strong solution $x_t$, $t>0$. 
This solution has finite moments of all order: 
for all $p\geq1$, $T<\infty$, and $x\in\NR^d$, there exists a constant
$C=C(x,p,T)<\infty$ such that for $0\leq t\leq T$,}
\qq
\NE_x(||x_t||^p)\leq C.
\label{bmoment}
\qqq

\noindent
When in need of emphasizing the dependence of
the solution to (\ref{seq}) on the initial condition $x$ and the
realization of the d-dimensional Brownian 
motion $w$ in the interval $[0,t]$, we shall write it as
$x_t(x,w([0,t]))$.  
We denote by $\CP^t$ the associated semigroup,
\qq
\CP^tf(x)=\NE_x(f(x_t))\equiv \int f(x_t(x,w([0,t])))\,{\bf dP}(w([0,t]),
\label{semi}
\qqq
where ${\bf P}$ is the $d$-dimensional Wiener measure,
by $\CA$ the generator of the semigroup, and by
$L$ the associated second order differential operator,
\qq
L=\sum_{i=1}^d X_{0}^i\,\partial_i+\sum_{i,j=1}^d a_{ij}\,\partial_i\partial_j\,,
\label{generator}
\qqq
where, with $\otimes$ denoting the tensor product,
\qq
a=\hf\sum_{k=1}^n X_k\otimes X_k\,.
\label{mata}
\qqq
From Assumption~3.1 on the process solution $x_t$ and the bounds
(\ref{xbo}) for the vector fields $X_l$, it follows that for each $t$
and $w[0,t]$, the map $x\mapsto x_t(x,w[0,t])$ is $\CC^\infty$ on 
$\NR^d$ with derivatives of all orders satisfying the stochastic
differential equation obtained from (\ref{seq}) by formal
differentiation. Furthermore,
for all multi-index $\alpha$, $p\geq1$, and $t\geq0$,
\qq
\NE(||\partial^\alpha x_t(x,\cdot)||^p)<\infty.
\label{bderivative}
\qqq
In the sequel, we will denote 
$U_t(x,w[0,t])=Dx_t(x,w[0,t])$, where $DX$ denotes the Jacobian matrix
of a vector field $X$ on $\NR^d$. 
The matrix $U_t$ is the linearized flow and 
it solves the equation, with initial condition $U_0={\bf 1}$,
\qq
dU_t=DX_0(x_t)U_t\,dt+\sum_{k=1}^n DX_k(x_t)U_t\,dw_k(t)\,.
\label{U_t}
\qqq
Below, $\NE_x U_t$ denotes 
$\int U_t(x,w[0,t])\,{\bf dP}(w[0,t])$.

Let us now assume the existence of an invariant probability measure
$\mu$ for the process solution $x_t$ of (\ref{seq})
and consider the covariance matrix at time $t$,
\qq
C_t(x)\equiv \NE_x(x_t\otimes x_t)-\NE_x\,x_t\otimes\NE_x\,x_t.
\label{cov}
\qqq
The following result is the starting point of the perturbative
analysis performed in subsequent sections. It provides an expression
for $\mu(C_t)$ in terms of the linearized flow $U_t$, where
$\mu(f)$ is a shorthand notation for $\int_{\NR^d}f(x)\,d\mu(x)$.

\claim{Proposition~3.2}
{\sl Suppose that the bounds (\ref{xbo}) and
Assumption~3.1 are satisfied. 
Suppose in addition that the invariant measure $\mu$ for the process 
solution $x_t$ of (\ref{seq}) is such that the functions
$x\mapsto\NE_x\,x_s^i$, 
$x\mapsto L\NE_x\,x_s^i$, 
and $x\mapsto a_{ij}(x)\NE_x\,U_s^{jl}$, 
belong to $L^2(\NR^d,d\mu)$ for all $i,j,l,$ and $s\leq t$.
Then,
\qq
\mu(C_t)=\int_0^t ds\sum_{k=1}^n \mu( \NE_. U_sX_k(.)\otimes \NE_. U_s X_k(.))\,.
\label{ineq}
\qqq
}

\medskip
\noindent{\sl Proof.} We will show below that the map
$s\mapsto\mu(\NE_. x_s\otimes \NE_. x_s)$ is 
differentiable, with
\qq
{d\over ds}\mu(\NE_. x_s\otimes \NE_. x_s)=
-\sum_{k=1}^n \mu( \NE_. U_sX_k(.)\otimes \NE_. U_s X_k(.))\,.
\label{P1}
\qqq
Identity (\ref{ineq}) thus follows from
the invariance of the measure $\mu$, since
\beqnarray
\mu(C_t)&=&\mu(\NE_.(x_t\otimes x_t))-
\mu(\NE_{.} x_t\otimes\NE_{.}x_t)
\\
&=&\mu(x\otimes x)-\mu(\NE_{.} x_t\otimes \NE_{.} x_t)
\\
&=&-\int_0^t ds\frac{d}{ds}\mu(\NE_. x_s\otimes \NE_. x_s).
\eeqnarray
To obtain (\ref{P1}), we first note that (\ref{bmoment}) implies that
any function $f\in\CC^2(\NR^d)$ with first derivatives of at most
polynomial growth is in the domain
of the generator $\CA$ with $\CA f=L f$.
Similarly, one easily checks that for such $f$,
(\ref{bderivative}) implies
$\CA(\CP_t f)=L(\CP_t f)$.
Therefore, Kolmogorov equation yields
${d\over ds}(\NE_x\,x_s\otimes \NE_x\,x_s)=
L\NE_x\,x_s\otimes \NE_x\,x_s+\NE_x\,x_s\otimes L\NE_x\,x_s$, which,
by H\"older inequality and our assumptions, belongs to $L^1(\NR^d,d\mu)$.
Thus,
\qq
{d\over ds}\mu(\NE_. x_s\otimes \NE_. x_s)
=\mu(L\NE_. x_s\otimes \NE_. x_s+\NE_. x_s\otimes L\NE_. x_s).
\label{P3}
\qqq
Let us next define for $f,g\in\CC^2(\NR^d)$,
\qq
\Gamma(f,g)\equiv L(fg)-fLg-gLf\,,
\qqq
which reads
\qq
\Gamma(f,g)=2\sum_{i,j=1}^d a_{ij}\,\partial_i f\,\partial_j g\,.
\label{P4}
\qqq
Since it follows from (\ref{bderivative}) that
$\partial_i\NE_x x_s^j=\NE_x U_s^{ji}$,
our assumptions imply as above that 
$\Gamma(\NE_.x_s^i,\NE_.x_s^j)\in L^1(\NR^d,d\mu)$ 
for all $i,j$. It follows in particular that
$L(\NE_. x_s\otimes\NE_. x_s)\in L^1(\NR^d,d\mu)$.
Because of the invariance of $\mu$ (which implies $\mu(Lf)=0$),
we are thus free to subtract from the $\mu$-expectation on the right
hand side of (\ref{P3}) a term 
$L(\NE_. x_s\otimes\NE_. x_s)$, so that
\qq
{d\over ds}\mu\Bigl((\NE_. x_s\otimes \NE_. x_s)_{ij}\Bigr)
=-\mu(\Gamma(\NE_. x_s^i\,,\NE_. x_s^j)).
\label{P31}
\qqq
Formula (\ref{P1}) finally follows from 
the computation, recalling (\ref{mata}),
\qq
\Gamma(\NE_. x_s^i,\NE_. x_s^j)(x)
=\sum_{k=1}^n\Bigl(\NE_x U_s X_k(x)\otimes\NE_x U_s X_k(x)\Bigr)_{ij}\,.
\qqq
This concludes the proof of Proposition~3.2.

Proposition 3.2 immediately implies the

\claim{Corollary 3.3.} {\sl Suppose that the hypothesis of
Proposition~3.2 are satisfied for all $t\geq0$.
Suppose in addition that
\qq
\lim_{t\rightarrow\infty}C_t=\mu(x\otimes x)-\mu(x)\otimes \mu(x)\equiv\Phi\,,
\qqq
in $L^1(\NR^d,d\mu)$.
Then,
\qq
\Phi=\int_0^\infty ds\sum_{k=1}^n \mu( \NE_. U_sX_k(.)\otimes \NE_. U_s X_k(.)).
\label{P6}
\qqq}

The expression (\ref{P6}) for the covariance matrix of a stationary state
is the basic formula that we shall use to
develop a perturbation expansion in the next section.  
Since both sides of (\ref{P6}) involve
an averaging with respect to $\mu$, it is not clear at first sight how
informations on $\mu$ can be extracted from (\ref{P6}). 
We observe, however, that in the case of a
linear drift $X_0$ and constant vector fields $X_k$, $k=1,\dots,n$,
all expectations may be dropped and
(\ref{P6}) becomes
\qq
\Phi_{\rm linear}=
\int_0^\infty ds\,U_s\Bigl(\sum_{k=1}^nX_k\otimes X_k\Bigr)
U_s^{\rm T}.
\label{P6linear}
\qqq
One thus recovers the standard formula for
the covariance of the stationary state of
a linear stochastic
equation with constant diffusion coefficients.
As we shall see in the next section, it is possible to iterate
this simple observation in order to begin a perturbation expansion.

Another feature of formula (\ref{ineq}) is to provide a link between
the covariance matrix $C_t$ and the
so-called Malliavin matrix.
The Malliavin matrix associated to equation (\ref{seq}) at time $t$
reads, in the normalization of \cite{Ikw},
\qq
M_t=\int_0^t ds\sum_{k=1}^n  U_tV_s X_k(x_s)\otimes U_tV_sX_k(x_s)\,,
\label{mal}
\qqq
where $V_s$ is the inverse matrix of $U_s$.
An easy computation reveals that $\mu(\NE_.M_t)$ can be expressed in a
form closely related to (\ref{ineq}), namely,
\qq
\mu(\NE_. M_t)=\int_0^t ds\sum_{k=1}^n \mu(\NE_. (U_s X_k(.)\otimes U_s X_k(.))).
\label{mal0}
\qqq
Indeed, we first observe that for $s\geq0$ fixed, $Y^t_s\equiv U_t V_s$ 
satisfies $Y^s_s={\bf 1}$ and
\qq
dY^t_s=DX_0(x_t)Y^t_s\, dt+ \sum_{k=1}^n DX_k(x_t)Y^t_s\, dw_k(t)
\label{eqy}
\qqq
for $t\geq s$.  
Comparing with (\ref{U_t}) yields that 
$Y_s^t=Y_s^t(x_s(x,w[0,s]),w[s,t])$
has the same $\NP$-distributions as 
$U_{t-s}(x_s(x,w[0,s]),\bar w[s,t])$,
where $\bar w(\tau)=w(\tau)-w(s)$ for $\tau\geq s$.
Furthermore, for $x$ fixed the map $w\mapsto Y_s^t(x,w[s,t])$ is
$w[0,s]$-independent.
Therefore, since $(x,w)\mapsto Y^t_s(x,w)X_k(x)\otimes Y^t_s(x,w)X_k(x)$
is measurable, one may use the Markov property of $x_t$ to write,
\qq
\NE_{x}(Y^t_s(x_s)X_k(x_s)\otimes Y^t_s(x_s)X_k(x_s))=
\NE_{x}(\NE_{y=x_s}(U_{t-s}(y)X_k(y)\otimes U_{t-s}(y)X_k(y))).
\label{eqyb}
\qqq
Identity (\ref{mal0}) then follows by using the invariance of the
measure $\mu$ and
changing variables in the integral over $s$ in (\ref{mal}).
As a consequence, Proposition~3.2
provides a lower bound on the
expectation of the Malliavin matrix.\footnote{The order relation is
  defined in the following way. For two matrices $X_1,X_2$, we say
  that $X_1\geq X_2$ whenever $X_1-X_2$ is a positive definite
  matrix.} 

\claim{Corollary 3.4.} {\sl One has
\qq
\mu(C_t)\leq\mu(\NE_.M_t).
\qqq
}
\medskip
\noindent{\sl Proof.}
The inequality simply follows from (\ref{ineq}), (\ref{mal0}), and
the matrix
\qq
\NE_x\Bigl[\Bigl(U_sX_k(x) -\NE_x U_sX_k(x)\Bigr)\otimes
\Bigl(U_sX_k(x) -\NE_x U_sX_k(x)\Bigr)\Bigr]
\label{mati}
\qqq
being positive definite.

\section{Perturbative analysis of the non-equilibrium anharmonic chain}
We shall analyze the effect of adding an anharmonic perturbation to a modification of
the model treated by Rieder, Lebowitz and Lieb \cite{rll}. We consider the case of a
harmonic chain with fixed ends to which one adds an anharmonic on-site potential, i.e.
in (\ref{Ham}), we set
\qq
U(x)=\hf\omega^2 x^2
\quad
{\rm and}
\quad
V=\hf\omega^2\kappa x^2+{1\over4}\lambda x^4.
\label{v_1}
\qqq
The model considered in \cite{rll} has $\kappa=0$ but the computation of the covariance
of the stationary state is very similar and the result is given below. We write the
equations of motions (\ref{eqq})-(\ref{eqp3}) under the matrix form,
\qq
\left( \begin{array}{cc}
d\un q\\
d\un p
\end{array}\right)=\Nb\left(\ba \un q\\\un p\ea\right)dt
-\lambda \left(\ba {\bf 0} \\ \CN(\un q) \ea\right)dt +
\left( \ba {\bf 0}\\ {\bf dw}\ea\right)
\label{eqla}
\qqq
with $\CN(\un q)$ and ${\bf dw}$ the
vectors in ${\bf R}^N$ given by $\CN_{i}(\un q)=q_i^3$ and
${\bf dw}_i=\delta_{1i}
\sqrt{2\gamma kT_1}\,dw_l+\delta_{Ni}\sqrt{2\gamma kT_N}\,dw_r$,
and
\qq
\Nb=\left(
\begin{array}{cc}
{\bf 0}& {\bf 1} \\-\Ng_{\kappa} & -\Na
\end{array}
\right)
\qqq
where
$\Ng_\kappa$ and $\Na$ are $N\times N$ matrices given by
$(\Ng_\kappa)_{ij}=\omega^2((2+\kappa)\delta_{ij}-\delta_{ij+1}-\delta_{ij-1})$ and
$\Na_{ij}=\gamma\delta_{ij}(\delta_{1j}+\delta_{Nj})$.
Above, ${\bf 1}$~denotes the unit matrix and ${\bf 0}$ the zero
matrix or vector, as is clear from the context. 
We note that the stochastic terms in (\ref{eqla}) 
are given by constant vector fields,
namely, in the notation of Section~3,
\qq
X_k=\left(\ba {\bf 0} \\ {\bf d}_k\ea\right)\quad
{\rm where}
\quad
({\bf d}_k)_j=\delta_{kj}\sqrt{2\gamma kT_k}\,,
\qqq
for $k=1,N$.
In particular, the coefficients $a_{ij}$ involved in the generator $L$
are constant. They are given by
\qq
\sum_{k=1,N}X_k\otimes X_k=
\left(\ba {\bf 0} & {\bf 0}\\ {\bf 0} & \Delta \ea\right)\,,
\label{int2.0}
\qqq
where $\Delta_{ij}=2\gamma k\delta_{ij}(T_1\delta_{1j}+T_N\delta_{Nj})$.
Furthermore, the linearized flow $U_t^\lambda$ 
of (\ref{eqla}) is given by
\qq
dU_t^\la=\Nb U_t^\lambda\,dt-3\lambda C^\lambda(t)U_t^\lambda\,dt\,,
\label{per2}
\qqq
where
\qq
C^\lambda(t)=\left(\ba {\bf 0} & {\bf 0} \\ \Nv^\lambda(t) & {\bf 0}\ea\right)\,,
\label{C}
\qqq
with
$\Nv^\lambda_{ij}(t)=\delta_{ij} q_i^2(t)$ and $q_i(t)$ the
$q_i$-component of the solution  of (\ref{eqla}) at time $t$.
Finally, we note that the matrix $\Nb$ in (\ref{eqla}) has the
property that all its eigenvalues have strictly negative real part.
A proof of this fact can be found in \cite{nak} modulo obvious
modifications. 

In order to study perturbatively the SNS
of our chain, we would like to use the identity (\ref{P6}).
However, some of the hypothesis of Corollary~3.3 related to the
invariant measure are not known to hold for equation
(\ref{eqla}) when $\lambda>0$. (The case $\lambda=0$ has been covered
in \cite{rll}.)
Although from a mathematical point of view, this is not a mere technical problem, but since
the main goal of this paper is to illustrate the use of formula
(\ref{P6}) for perturbative analysis on a specific example, we will
assume that these hypothesis hold, see Assumption~4.1 below and the
remark that follows. On the other hand, Assumption~3.1, i.e., the
existence of strong solutions and their moments, follows from
standard techniques and we briefly discuss it now.
We first note that for $\lambda>0$, the
function $\tilde H(\un q,\un p)=2N+H(\un q,\un p)$, with $H$ the
Hamiltonian given by (\ref{Ham}) and (\ref{v_1}), satisfies
\qq
\tilde H(\un q,\un p)\geq C(1+||\un q||^2+||\un p||^2),
\qqq
for some $C>0$ and all $(\un q,\un p)\in\NR^{2N}$.
Thus, $\tilde H$ is a $\CC^2(\NR^{2N})$ confining function.
Furthermore, one computes
\qq
(L\tilde H)(\un q,\un p)=
-\gamma(p_1^2+p_N^2)+2\gamma k(T_1+T_N),
\label{LHtil}
\qqq
which implies that $L\tilde H$ is uniformly bounded by above.
A classical result, see e.g. \cite{has}, Thm~4.1, then ensures for all initial
conditions $(\un q,\un p)\in\NR^{2N}$ the
existence of a unique global strong solution to (\ref{eqla}).
Regarding the 
bounds (\ref{bmoment}), they are an immediate consequence of
the following a priori bound. For any
$\theta\leq(2k\max\{T_1,T_N\})^{-1}$, one has
\qq
\NE_{(\un q,\un p)}\Bigl[e^{\theta H(\un q_t,\un p_t)}\Bigr]\leq
e^{2\gamma k\theta(T_1+T_N)t}\,e^{\theta H(\un q,\un p)}\,.
\label{beH}
\qqq
Bound (\ref{beH}) can be obtained in a similar way as in the proof 
of Lemma~3.5 in~\cite{rbt}.
However, the existence of a unique invariant measure for (\ref{eqla})
is still an open problem.  We thus introduce the
following

\claim{Assumption 4.1.} {\sl The finite time truncated two-point
correlation function of the process defined by (\ref{eqla}) converges
to the covariance matrix of a unique stationary measure $\mu^\la$ in
$L^1(\NR^{2N},d\mu^\la)$-norm.
Furthermore, the decay properties of $\mu^\lambda$ are such that
$\NE_{(\un q,\un p)}[(\un q_t,\un p_t)]$, 
$L\NE_{(\un q,\un p)}[(\un q_t,\un p_t)]$, and 
$\NE_{(\un q,\un p)}[U^\lambda_t]$ belong to $L^2(\NR^{2N},d\mu^\lambda)$.
}
\medskip

\claim{Remark.}  The uniqueness of the invariant measure is proved in
\cite{eck3,rbt} for a large class of anharmonic chains. The invariant
measure has a smooth density with exponential decay and is shown to be
mixing\,\footnote{In \cite{rbt}, the result is actually stronger. The
convergence to the unique invariant measure is shown to be
exponential.}.  An important restriction is that the potential $U$
must not grow asymptotically slower than $V$, and thus equation 
(\ref{eqla}) does
not fall into the class covered in \cite{eck3,rbt}. However, as is
argued in \cite{rbt}, the fact that the on-site potential grows faster
than the nearest-neighbour interaction should not affect the
ergodic properties of the measure but only the rate
of convergence. 
Although we could consider a similar anharmonic chain with an
additional quartic term in the nearest-neighbour interaction,
the equations that one then needs to solve, see below, are
computationally more involved.
Furthermore, restricting to (\ref{eqla}) will allow us to compare our 
results to the
usual $\lambda\phi^4$ expansion when the temperatures of the two baths
are equal.

\medskip
Provided Assumption~4.1 holds,
let $\Phi^\lambda$ denote the covariance matrix of the unique
stationary state of equation (\ref{eqla}) and express it according to
(\ref{P6}) as
\qq
\Phi^\la=\int_0^\infty dt\sum_{k=1,N} \mu^\la
( \NE_. U^\la_t X_k\otimes \NE_. U^\la_t X_k)\,.
\label{per1}
\qqq
We first briefly review the harmonic case $\lambda=0$.
As mentioned at the end of the previous
section,
one obtains from (\ref{per1})
\qq
\Phi^0=\int_0^\infty dt\,e^{{\bf b}t}\,\ND\,e^{{\bf b}^{\rm T}t}\,,
\label{Phi01}
\qqq
where
\qq
\ND=\sum_{k=1,N}X_k\otimes X_k=
\left(\ba {\bf 0} & {\bf 0}\\ {\bf 0} & \Delta \ea\right)\,,
\label{int2}
\qqq
with $\Delta_{ij}=2\gamma k\delta_{ij}(T_1\delta_{1j}+T_N\delta_{Nj})$.
Since the eigenvalues of $\Nb$ have strictly negative
real part, the integral in (\ref{Phi01}) is convergent
and it follows from integrating by parts in $\Nb\Phi^0$
that $\Phi^0$ must satisfy the equation
\qq
\Nb\Phi^0+\Phi^0\Nb^{\rm T}=-\ND\,.
\label{int1}
\qqq
The unique solution of this equation has been explicitly derived in
\cite{rll}. It is given by
\qq
\Phi^0=\left(\ba \Phi_x^0 & \Phi_z^0\\ -\Phi_z^0 &\Phi_y^0 \ea\right)
\label{Phi03}
\qqq
where,
denoting $T=\frac{T_1+T_N}{2}$, $\eta=\frac{T_1-T_N}{2T}$,
and $\NG_\kappa=\omega^{-2}\Ng_\kappa$,
\beqnarray
\Phi_x^0&=&\frac{kT}{\omega^2}(\NG_\kappa^{-1}+\eta\NX^0),
\label{x}
\\
\Phi_y^0&=&kT({\bf 1}+\eta \NY^0),
\label{y}
\\
\Phi_z^0&=&{kT\over\gamma}\eta\NZ^0,
\label{z}
\eeqnarray
and
\def\adots{\mathinner{\mkern2mu\raise1pt\hbox{.}
\mkern3mu\raise4pt\hbox{.}\mkern1mu\raise7pt\hbox{.}}}
\myeqnarray{15}
\NX^0&=&\pmatrix{
\phi_1&\phi_2&          &\phi_{N-2}&\phi_{N-1} & 0\cr
\phi_2    &\adots   &\adots   &\adots    &     \adots        &-\phi_{N-1}\cr
\phi_3    &\adots   &\adots   &\adots    &\adots       &             \cr
              &\adots   &\adots   &\adots    &\adots       &             \cr
              &         &\adots   &\adots    &\adots       &           \cr
\phi_{N-1} &         &         &\adots    &\adots       &-\phi_2    \cr
0&  -\phi_{N-1}       & &         &-\phi_2   &-\phi_1            \cr
},
\label{matrixX0}
\\
\NY^0_{ij}&=&\delta_{ij}(\delta_{i1}-\delta_{iN})-\nu\NX^0_{ij}\,,
\label{NY}
\\
\NZ^0&=&\pmatrix{
0             &\phi_1&\phi_2&          &\phi_{N-2}&\phi_{N-1}\cr
-\phi_1    &\ddots   &\ddots   &\ddots    &             &\phi_{N-2}\cr
-\phi_2    &\ddots   &\ddots   &\ddots    &\ddots       &             \cr
              &\ddots   &\ddots   &\ddots    &\ddots       &             \cr
              &         &\ddots   &\ddots    &\ddots       &\phi_2    \cr
              &         &         &\ddots    &\ddots       &\phi_1    \cr
-\phi_{N-1}&         &         &-\phi_2 &-\phi_1   & 0            \cr
}.
\label{matrixZ0}
\eeqnarray
Above, $\nu=\frac{\omega^2}{\gamma^2}$ and the quantities $\phi_j$,
$1\leq j\leq N-1$, satisfy the equation
\qq
\sum_{j=1}^{N-1}(\NG_{\nu+\kappa}^{(N-1)})_{ij}\phi_j=\delta_{1i}\,,
\label{equphi}
\qqq
where $\NG_{\nu+\kappa}^{(k)}$ denotes the $k$-square matrix given by
$(\NG^{(k)}_{\nu+\kappa})_{ij}=(2+\nu+\kappa)\delta_{ij}-\delta_{i,j+1}-\delta_{i,j-1}$.
The solution of (\ref{equphi}) is given by
\qq
\phi_j=\frac{\sinh(N-j)\alpha}{\sinh N\alpha}\,,
\label{phi}
\qqq
with $\alpha$ defined by $\cosh\alpha=1+(\nu+\kappa)/2$.
Hence, one has for large $N$ and fixed $j$ the asymptotic formula
$\phi_j=e^{-\alpha j}$.  In the context of SNS, one usually defines
the temperature to be the average kinetic energy, 
i.e. in our case,
\qq
T_i=(\Phi^0_y)_{ii}.
\label{temp}
\qqq
It is easy to see that the above solution yields an exponentially
flat profile in the bulk of the chain.

We now turn to the first-order perturbation of the anharmonic chain.
We first introduce our second assumption on the process solution of (\ref{eqla}).

\claim{Assumption A2.} {\sl The measure $\mu^\la$ is
absolutely continuous with respect to the Lebesgue measure and as
a function of $\la$ its density $\rho^\la(x)$ is $ C^\infty$ in a
neighbourhood of $0$.  For all $x$, all derivatives are bounded in
a neighbourhood of $0$. }

\claim{Remark.} The proof of this fact should follow from an analysis similar
to the ones developed in \cite{eck4} or \cite{wata} to prove the smoothness of the
probability transitions in a parameter of the related stochastic differential
equations.

\medskip
To derive an expression for 
$\Phi^1\equiv\frac{d}{d\la}\Phi^\la|_{\la=0}$, we compute from (\ref{per1})
\beqnarray
\Phi^1&=&\frac{d}{d\la}\Phi^\la|_{\la=0}
\\
&=&\mu^1\Bigl(\int_0^\infty dt\sum_{i=1,N} \NE_. U^0_tX_i(.)
\otimes \NE_. U^0_t X_i(.)\Bigr)\nonumber
\\
&&\qquad+\ \mu^0\Bigl(\int_0^\infty dt
\sum_{i=1,N}\NE_.\frac{d}{d\la} U^\la_t|_{\la=0}X_i(.)
\otimes \NE_. U^0_t X_i(.)\Bigr)+{\rm tr.}\,,
\label{Phi1}
\eeqnarray
and observe that the first term vanishes because
$\mu^1\equiv\frac{d}{d\la}\mu^\la|_{\la=0} $ integrates constants to zero.
In order to compute the last terms, we first
evaluate $W_t\equiv\frac{d}{d\la} U^\la_t|_{\la=0}\,$.
Deriving with respect to $\la$ on both sides of equation (\ref{per2}), we get
\qq
dW_t= \Nb\,W_t\,dt-3\,C^0(t)\,U^0_t\,dt,
\label{V1}
\qqq
from which it follows that, since $W_0=0$,
\qq
W_t=-3\int_0^t ds\,e^{\Nb(t-s)}C^0(s)\,e^{\Nb s}\,.
\label{V2}
\qqq
Inserting (\ref{V2}) in (\ref{Phi1}), we obtain, using in addition
the invariance of $\mu^0$,
\beqnarray
\Phi^1&=&-3\int_0^\infty dt\int_0^t ds\sum_{i=1,N}e^{\Nb(t-s)}\NN\,e^{\Nb s}
X_i\otimes e^{\Nb t}X_i +{\rm tr.}\,,
\\
&=&-3\int_0^\infty dt\int_0^t ds\ e^{\Nb(t-s)}\,\NN\,e^{\Nb s}\,
\ND\,e^{\Nb^{\rm T} t}+{\rm tr.}\,,
\label{Phi11}
\eeqnarray
where $\ND$ is given by (\ref{int2}) and
\qq
\NN=\mu^0(C^0(0))=
\left(\ba 0 & 0 \\ {\rm diag}(\Phi^0_x) & 0\ea\right).
\label{NNexpr}
\qqq
Exchanging the integrations over $t$ and $s$ and changing variables
leads to
\qq
\Phi^1=-3\int_0^\infty dt\,e^{\Nb t}\NN
\Bigl(\int_0^\infty ds\,e^{\Nb s}\,\ND\,e^{\Nb^{\rm T}s}\Bigr)
e^{\Nb^{\rm T}t}+{\rm tr.}\,,
\qqq
which, with (\ref{Phi01}), finally yields,
\qq
\Phi^1=-3\int_0^\infty dt\,e^{\Nb t}
(\NN\Phi^0+\Phi^0\NN^{\rm T})e^{\Nb^{\rm T}t}\,.
\label{Phi10}
\qqq

The method used to derive the above equation will also provide the equations for the
next orders of the perturbative expansion.  However, obtaining them concretely requires
some more work and we reserve that part and the general Feynman rules for a further
publication. 
We note that integrating by parts in (\ref{Phi10}) yields the
equation for $\Phi^1$
\qq
\Nb\Phi^1+\Phi^1\Nb^{\rm T}=3(\NN\Phi^0+\Phi^0\NN^{\rm T}).
\label{Phi12}
\qqq
In Section~6, we will
derive an explicit expression for $\Phi^1$ and thus for the first
order correction to the heat current and temperature profile.
It turns out to be easier to do so by solving equation (\ref{Phi12})
rather than by using (\ref{Phi10}).
In the next section, we first make a few preliminary remarks
about equations of the form (\ref{Phi12}).

\section{Solving the equation for the first order}

The symmetry properties of the inhomogeneous term in equation
(\ref{Phi12}) will play a special role. We will need to consider
symmetry properties both with respect to the diagonal and to the
cross-diagonal.

\claim{Notation.} For a $K$-square matrix $\NM$, we denote by
$\NM^{\rm C}$ the transpose of $\NM$ with respect to the
cross-diagonal, namely, $(\NM^{\rm C})_{ij}=\NM_{K+1-j,K+1-i}\,$.

\claim{Definition.} {\sl We call a square matrix $\NM$
c-symmetric or c-antisymmetric if
$\NM^{\rm C}=\NM$ or, respectively, $\NM^{\rm C}=-\NM$.
Denoting
\qq
\NJ=\left(\begin{array}{cc} {\bf 0} & {\bf 1} \\ {\bf 1} & {\bf 0} \end{array}\right),
\label{defJ}
\qqq
we call a $2N$-square matrix $\NM$ CT-symmetric or CT-antisymmetric
if $\NM^{\rm C}=\NJ\NM\NJ$ or, respectively,
$\NM^{\rm C}=-\NJ\NM\NJ$.}

\medskip
\noindent

We first list a few properties of equations of the form (\ref{Phi12}).

\claim{Lemma~5.1} {\sl
Let $\Nb$ as above and $\NH$ a $2N$-square matrix.
One has:
\renewcommand{\theenumi}{\alph{enumi}}
\renewcommand{\labelenumi}{(\theenumi).}
\begin{enumerate}
\item The unique solution of the equation
\qq
\Nb\Phi+\Phi\Nb^{\rm T}=\NH
\label{lemeq}
\qqq
is given by
\qq
\Phi=-\int_0^\infty dt\,e^{\Nb t}\,\NH\,e^{\Nb^{\rm T}t}.
\label{lemsol}
\qqq
\item
If $\NH$ is CT-symmetric or CT-antisymmetric, then
$\Phi$ is CT-symmetric or, respectively, CT-antisymmetric.
\item If $\NH$ is of the form
\qq
\NH=\left(\begin{array}{cc} {\bf 0} & * \\ \ast & * \end{array}\right),
\label{specform}
\qqq
then the solution of (\ref{lemeq}) is of the form
\qq
\Phi=\left(\begin{array}{cc} \NX & \NZ \\ -\NZ & \NY \end{array}\right).
\label{lemform}
\qqq
\end{enumerate}
}
\medskip
\noindent{\sl Proof.} Point {(a)} follows from the
matrix $\Nb$ having all its eigenvalues with strictly negative real part.
Indeed, this property implies that the operator 
$\Phi\mapsto\Nb\Phi+\Phi\Nb^{\rm T}$ is invertible, and integrating by
part in $\Nb\Phi$ reveals that (\ref{lemsol}) is the unique solution
of (\ref{lemeq}). 
Point (c) is obvious, whereas (b) follows from the identity 
$\NJ\Nb^{\rm C}\NJ=\Nb^{\rm T}$ and uniqueness of the solution
of (\ref{lemeq}).

\medskip
Lemma~5.1 implies in particular that $\Phi^1$ is the unique solution of
(\ref{Phi12}) and is of the form
\qq
\Phi^1=\left(\ba \Phi^1_x &\Phi^1_z \\ -\Phi^1_z & \Phi^1_y \ea\right).
\label{Phi1formg}
\qqq
In particular, it follows from (\ref{Phi1formg}) and $\Phi^1$ being
symmetric that $\Phi^1_z$ is antisymmetric.
In order to find an expression for the solution of equation
(\ref{Phi12}),
we decompose the inhomogeneous term on the RHS of (\ref{Phi12})
into powers of $\eta$ and solve the equation
separately for each case.
One has
\qq
3(\NN\Phi^0+\Phi^0\NN^{\rm T})={3k^2T^2\over\omega^4}(\NH_0+\eta\NH_1+\eta^2\NH_2),
\qqq
where, cf.~(\ref{Phi03})-(\ref{z}) and (\ref{NNexpr}),
\myeqnarray{10}
\NH_0&=&\left(\begin{array}{cc}
{\bf 0} & \NG_\kappa^{-1}\bar\NV_0 \\ \bar\NV_0\NG_\kappa^{-1} & {\bf 0}
\end{array}\right),
\\
\NH_1&=&\left(\begin{array}{cc}
{\bf 0} & \NX^0\bar\NV_0+\NG_\kappa^{-1}\bar\NV_1 \\
\bar\NV_1\NG_\kappa^{-1}+\bar\NV_0\NX^0 & \gamma\nu[\bar\NV_0,\NZ^0]
\end{array}\right),
\\
\NH_2&=&\left(\begin{array}{cc}
{\bf 0} & \NX^0\bar\NV_1 \\ \bar\NV_1\NX^0 & \gamma\nu[\bar\NV_1,\NZ^0]
\end{array}\right),
\eeqnarray
with
\qq
\bar\NV_0\equiv{\rm diag}(\NG_\kappa^{-1}),
\quad \bar\NV_1\equiv{\rm diag}(\NX^0).
\qqq
In the sequel, we will denote $(\bar\NV_0)_{ij}=\delta_{ij}g_i$, where
$g_i=(\NG_\kappa^{-1})_{ii}$ read
\qq
g_i={\sinh i\bar\alpha\over\sinh\bar\alpha}{\sinh(N+1-i)\bar\alpha\over\sinh
(N+1)\bar\alpha}\,,
\label{gi}
\qqq
with $\bar\alpha$ defined by $\cosh\bar\alpha=1+\kappa/2$.
Writing
\qq
\Phi^1={3k^2T^2\over\omega^4}(\Phi^1_0+\eta\Phi^1_1+\eta^2\Phi^1_2),
\label{decompPhi}
\qqq
one thus obtains that $\Phi^1_l$, $l=0,1,2$, is the unique solution of
\qq
\Nb\Phi^1_l+\Phi^1_l\Nb^{\rm T}=\NH_l\,.
\label{equPhil}
\qqq
In order to scale out the constants in $\Nb$, we denote
for $l=0,1,2$,
\qq
\Phi^1_l=\left(\begin{array}{cc}
{1\over\omega^2} \NX_l & {1\over\gamma}\NZ_l \\ -{1\over\gamma}\NZ_l & \NY_l
\end{array}\right),
\label{Philansatz}
\qqq
together with
\qq
\NR=\gamma^{-1}\Na\,,\quad\NG_\kappa=\omega^{-2}\Ng_\kappa,
\label{NRNG}
\qqq
namely, $\NR_{ij}=\delta_{ij}(\delta_{1j}+\delta_{Nj})$ and
$(\NG_\kappa)_{ij}=(2+\kappa)\delta_{ij}-\delta_{ij+1}-\delta_{ij-1}$.
The zero order term in (\ref{decompPhi})
is just the first-order perturbation of
the anharmonic chain at the equilibrium $T_1=T_N$.
Inserting (\ref{Philansatz}) into (\ref{equPhil}) for $l=0$ yields the
equivalent system of equations for $\NX_0,\NY_0$ and $\NZ_0$
\beqnarray
\NY_0&=&\NX_0\NG_\kappa+\NZ_0\NR+\NG_\kappa^{-1}\bar\NV_0,
\label{zero2}
\\
\lbrack\NG_\kappa,\NZ_0]&=&-{1\over\nu}\{\NR,\NY_0\},
\label{zero3}
\eeqnarray
with the requirement that $\NX_0,\NY_0$ are symmetric and $\NZ_0$ is
antisymmetric.
One easily checks that its unique solution is given by
\begin{equation}
\NX_0=-\NG_\kappa^{-1}\bar\NV_0\NG_\kappa^{-1}\,,\quad
\NY_0=0\,,\quad
\NZ_0=0\,,
\label{solzero}
\end{equation}
thus recovering, as expected, the first-order correction of
the $\lambda\phi^4$ model.
Proceeding similarly for $\Phi^1_1$ and $\Phi^1_2$, one finds that
$\NX_1,\NY_1,\NZ_1$ solve
\beqnarray
\NY_1&=&\NX_1\NG_\kappa+\NZ_1\NR+(\NX^0\bar\NV_0+\NG_\kappa^{-1}\bar\NV_1),
\label{lin2}
\\
\lbrack\NG_\kappa,\NZ_1]&=&-{1\over\nu}\{\NR,\NY_1\}+[\NZ^0,\bar\NV_0],
\label{lin3}
\eeqnarray
whereas $\NX_2,\NY_2,\NZ_2$ solve
\beqnarray
\NY_2&=&\NX_2\NG_\kappa+\NZ_2\NR+\NX^0\bar\NV_1,
\label{qua2}
\\
\lbrack\NG_\kappa,\NZ_2]&=&-{1\over\nu}\{\NR,\NY_2\}+[\NZ^0,\bar\NV_1].
\label{qua3}
\eeqnarray
Furthermore, using the c-symmetry properties of the solution
$\NX^0$ and $\NZ^0$ of the
harmonic case, cf.~(\ref{matrixX0}) and (\ref{matrixZ0}),
one easily checks that $\NH_1$ is CT-antisymmetric, whereas $\NH_2$
is CT-symmetric. This implies that
$\NX_1,\NY_1$ are c-antisymmetric and $\NZ_1$ is c-symmetric,
whereas $\NX_2,\NY_2$ are c-symmetric and $\NZ_2$ is c-antisymmetric.
This simply reflects the fact that changing the sign of $\eta$
corresponds to interchanging the reservoirs at the ends of the chain.

In the next section, we will derive explicit expressions for the solutions of the above
equations. 
To this end, we will need the following identities. 
Let $\NX$ be a solution of
\qq
[\NG_\kappa,\NX]=\CU,
\label{GXU}
\qqq
with $\CU$ a given matrix.
It thus follows from
$[\NG_\kappa,\NX]_{ij}=\CU_{ij}$ that
\qq
\NX_{i,\,j+1}-\NX_{i-1,\,j}=\CU_{ij}+(\NX_{i+1,\,j}-\NX_{i,\,j-1}),
\label{inteform0}
\qqq
where matrix elements with an index equals to zero or $N+1$
are set to zero.
Let us first consider $\NX$ antisymmetric.
In particular, $\NX$ is entirely determined by its elements $\NX_{ij}$
with $i<j$ and satisfies
$\NX_{j+1,\,i}-\NX_{j,\,i-1}=-(\NX_{i,\,j+1}-\NX_{i-1,\,j})$.
For $i\leq j$, applying (\ref{inteform0}) recursively $j-i$ times
thus leads to
\qq
\NX_{i,\,j+1}-\NX_{i-1,\,j}={1\over2}\sum_{l=0}^{j-i}\CU_{i+l,\,j-l}\,.
\label{inteform}
\qqq
This gives all matrix elements $\NX_{1j}$, $1< j\leq N$.
Applying (\ref{inteform})
recursively $i-1$ times finally leads to
\qq
\NX_{ij}={1\over2}\sum_{k=0}^{i-1}\sum_{l=0}^{j-i-1}\CU_{i+l-k,\,j-l-k-1}\,,
\label{eforms}
\qqq
for $i,j$ such that $i<j$.
Proceeding similarly, one obtains for a c-antisymmetric matrix $\NX$
satisfying (\ref{GXU}),
\qq
\NX_{ij}={1\over2}\sum_{k=0}^{i-1}\sum_{l=0}^{N-i-j}\CU_{i+l-k,\,j+l+k+1}\,,
\label{eformcs}
\qqq
for $i+j\leq N$.
If $\NX$ is both antisymmetric and c-antisymmetric, one iterates 
identity (\ref{inteform}) 
$N+1-i-j$ times
to obtain
\qq
\NX_{ij}=-{1\over4}\sum_{k=0}^{j-i-1}\sum_{l=0}^{N-i-j}
\CU_{i+l+k+1,\,j+l-k}\,,
\label{eformscs}
\qqq
for $i<j$ and $i+j\leq N$.
Finally, proceeding similarly but without assuming any symmetry
properties, one derives an expression for $\NX$ depending both on
$\CU$ and the first line of $\NX$,
\qq
\NX_{ij}=\sum_{k=1}^i\NX_{1,i+j-2k+1}
-\sum_{k=1}^{i-1}\sum_{l=1}^{i-k}\CU_{i+1-k-l,\,j-k+l}\,,
\label{eformgen}
\qqq
for $1<i\leq j$ and $i+j\leq N+1$.
Formula (\ref{eformgen}) will be used later for $\NX$ symmetric and
c-symmetric. It reflects the fact that in such cases, the solution of (\ref{GXU})
is determined up to a
polynomial $P(\NG)$, that is up to $N$ independent variables which can
be supplemented as the first line of $\NX$.

\section{The first-order correction}
In this section, we derive an expression for the first-order correction to the heat
current and temperature profile.
We find that the part corresponding to the heat current is uniformly bounded in $N$.
In particular, a first-order perturbation does not reveal any sign that
Fourier law might hold in such anharmonic models, as numerical studies
indicate, see e.g.~\cite{huli}.
Indeed, if Fourier law holds whenever $\lambda$ is finite,
one might
expect the derivatives of the heat current to develop a singularity at $\lambda=0$ when
$N\rightarrow\infty$.

Regarding the temperature profile,
the part of the solution proportional to $\eta$ is exponentially
decaying in the bulk of the chain whenever $\kappa>0$.
The decay rate is slower than in the purely harmonic case.
For $\kappa=0$, the profile proportional to $\eta$ is
linear in the bulk of the chain and
we compute its slope explicitly.
However as
explained in the introduction, the sign is ``wrong", in the sense that the
linear profile has the {\sl lowest} temperature close to the
{\sl hottest} bath
and the {\sl highest} temperature close to the {\sl coldest} bath.
The same type of phenomenon is 
present for $\kappa>0$, see Figure~1.
Moreover, we
observe that the part proportional to $\eta^2$ gives a significant contribution, which
results in a shift of the temperature at  the middle point  of the chain. The
temperature at this point is no more the arithmetic mean of the baths temperatures.
Although surprising, this is a phenomenon which seems to be observed in numerical studies of certain anharmonic
chains, see~\cite{huli}.

\subsection{First-order correction to the heat current}

In our model, the heat current in the SNS is given by
$(\Phi_z^\lambda)_{i,i+1}$. The first-order correction will thus be
given in terms of, cf.~(\ref{decompPhi}) and (\ref{Philansatz}),
\qq
\Phi_z^1={3k^2T^2\over\gamma\omega^4}(\NZ_0+\eta\NZ_1+\eta^2\NZ_2).
\label{decphi2}
\qqq
By (\ref{solzero}), $\NZ_0$ does not contribute and one easily checks that
for $1\leq i\leq N-1$,
\qq
(\NZ_2)_{i,i+1}=0.
\label{cur2}
\qqq
That is, $\NZ_2$ does not contribute to the current either.
Indeed, recall that $\NZ_2$ is antisymmetric and satisfies equation
(\ref{qua3}).
Since $\{\NR,\NY_2\}$ is a
bordered matrix and $[\NZ^0,\bar\NV_1 ]$ is zero on the diagonal,
one obtains by using formula (\ref{eforms}) that
\qq
-{1\over\nu}(\NY_2)_{11}=(\NZ_{2})_{12}=(\NZ_{2})_{23}=\ldots=(\NZ_{2})_{N-1,N}\,.
\qqq
On the other hand, the c-antisymmetry of $\NZ_2$ implies that
$(\NZ_{2})_{12}=-(\NZ_{2})_{N-1,N}$,
which leads to (\ref{cur2}). We note for later use that this also
implies
\qq
(\NY_2)_{11}=0.
\label{Y211zero}
\qqq
It thus remains to consider the contribution of $\NZ_1$.
Since $\NZ_1$ is antisymmetric, one obtains from (\ref{lin3})
that
\qq
\NZ_1=\NZ+\CZ,
\label{decZ}
\qqq
where $\NZ$ and $\CZ$ are given by formula (\ref{eforms}) with $\CU$
replaced by $-{1\over\nu}\{\NR,\NY_1\}$ and,
respectively, $[\NZ^0,\bar\NV_0]$.
We first observe that $\{\NR,\NY_1\}$ is a bordered symmetric matrix, so that
formula (\ref{eforms}) yields
\qq
\NZ=\pmatrix{
0             &\varphi_1&\varphi_2&          &\varphi_{N-2}&\varphi_{N-1}\cr
-\varphi_1    &\ddots   &\ddots   &\ddots    &             &\varphi_{N-2}\cr
-\varphi_2    &\ddots   &\ddots   &\ddots    &\ddots       &             \cr
              &\ddots   &\ddots   &\ddots    &\ddots       &             \cr
              &         &\ddots   &\ddots    &\ddots       &\varphi_2    \cr
              &         &         &\ddots    &\ddots       &\varphi_1    \cr
-\varphi_{N-1}&         &         &-\varphi_2&-\varphi_1   &0            \cr
},
\label{matrixZ}
\qqq
where the quantities $\varphi_1,\dots,\varphi_{N-1}$ are related to the
first line of $\NY_1$, namely, for $j=1,\dots,N-1$,
\qq
\quad\nu\varphi_j=-(\NY_1)_{1j}\,.
\label{phijY}
\qqq
Furthermore, $[\NZ^0,\bar\NV_0]$ having zero diagonal implies
that $\CZ_{i,i+1}=0$.
One therefore obtains
\qq
(\NZ_1)_{i,i+1}=\NZ_{i,i+1}=\varphi_1\,.
\label{cur1}
\qqq
In order to compute the vector $\varphi\in{\bf R}^{N-1}$,
one considers the first line of equation (\ref{lin2}) for
$\NY_1$ into which one substitutes identity (\ref{phijY}).
We first need to compute $\NX_1$.
Equation (\ref{lin2}) and the symmetry properties of
$\NX_1,\NY_1$ and $\NZ_1$ imply that $\NX_1$ satisfies
\beqnarray
[\NG_\kappa,\NX_1]&=&\{\NR,\NZ_1\}+([\NX^0,\bar\NV_0]
+[\NG_\kappa^{-1},\bar\NV_1])
\\
&=&\{\NR,\NZ\}+\{\NR,\CZ\}+([\NX^0,\bar\NV_0]+[\NG_\kappa^{-1},\bar\NV_1]).
\label{lin1}
\eeqnarray
Since $\NX_1$ is c-antisymmetric, it follows from (\ref{lin1}) that
\qq
\NX_1=\NX+\CX,
\label{decX}
\qqq
where $\NX$ and $\CX$ are given by formula (\ref{eformcs}) with $\CU$
replaced by $\{\NR,\NZ\}$ and, respectively,
$\{\NR,\CZ\}+([\NX^0,\bar\NV_0]+[\NG_\kappa^{-1},\bar\NV_1])$.
Using that $\{\NR,\NZ\}$ is a bordered antisymmetric matrix, one obtains from
(\ref{eformcs}) and (\ref{matrixZ}) that
\qq
\NX=\pmatrix{
\varphi_1&\varphi_2&          &\varphi_{N-2}&\varphi_{N-1} & 0\cr
\varphi_2    &\adots   &\adots   &\adots    &     \adots        &-\varphi_{N-1}\cr
\varphi_3    &\adots   &\adots   &\adots    &\adots       &             \cr
              &\adots   &\adots   &\adots    &\adots       &             \cr
              &         &\adots   &\adots    &\adots       &           \cr
\varphi_{N-1} &         &         &\adots    &\adots       &-\varphi_2    \cr
0&  -\varphi_{N-1}       & &         &-\varphi_2   &-\varphi_1            \cr
}.
\label{matrixX}
\qqq
Equation (\ref{lin2}) now reads
\qq
\NY_1=\NX\NG_\kappa+\NZ\NR+\NW,
\label{lin22}
\qqq
with
\qq
\NW=\CX\NG_\kappa+\CZ\NR+(\NX^0\bar\NV_0+\NG_\kappa^{-1}\bar\NV_1),
\label{inhomY2}
\qqq
and since $(\NX\NG_\kappa+\NZ\NR)_{1j}=(\NG_\kappa\NX_{1\cdot})_j=(\NG_\kappa^{(N-1)}\varphi)_j$
for $j=1,\dots,N-1$, where $\NG_\kappa^{(k)}$ denotes the
$k$-square version of $\NG_\kappa$, it follows from (\ref{phijY}) that
\qq
\NG^{(N-1)}_{\nu+\kappa}\varphi=-\Nw\,,
\label{eqphij}
\qqq
where 
$\Nw\in\NR^{N-1}$ is given by $\Nw_j=\NW_{1j}$, $j=1,\dots,N-1$.
Therefore, one finally obtains,
recalling that $\eta={T_1-T_N\over2T}\,$,
\qq
(\Phi_z^1)_{i,i+1}={3k^2T(T_1-T_N)\over2\gamma\omega^4}\,\varphi_1\,,
\qqq
with $\varphi$ given by $\varphi=-[\NG^{(N-1)}_{\nu+\kappa}]^{-1}\Nw$.
As $(\Phi_z^1)_{i,i+1}$ represent the first-order correction to the current, it is
consistent to see that they are all equal to each other.

Before turning to the first-order correction of the temperature
profile, we study the behaviour of $\varphi_1$ with $N$. 
We first note that $\NX$ solves the equation
$[\NG_\kappa,\NX]=\{\NR,\NZ\}$, as is easily checked from (\ref{matrixZ}) and
(\ref{matrixX}).
This implies that $\CX$ solves, cf.~(\ref{lin1}) and (\ref{decX}),
\qq
[\NG_\kappa,\CX]=\{\NR,\NZ\}+([\NX^0,\bar\NV_0]+[\NG_\kappa^{-1},\bar\NV_1]),
\qqq
which in turn implies, by using in
addition the symmetry properties of the matrices involved in
(\ref{inhomY2}), that $\NW$ is c-antisymmetric and
satisfies the equation
\qq
[\NG_\kappa,\NW]=\NG_\kappa\CZ\NR+\NR\CZ\NG_\kappa+(\NG_\kappa\NX^0\bar\NV^0-\bar\NV^0\NX^0\NG_\kappa).
\label{eqU}
\qqq
Hence, $\NW_{1N}=0$ and it follows from formula (\ref{eformcs})
that
\qq
\Nw=\Nw^{(1)}+\Nw^{(2)}\,,
\label{solU}
\qqq
where, for $1\leq j\leq N-1$,
\beqnarray
\Nw^{(1)}_j&=&{1\over2}\sum_{l=1}^{N-j}(\NG_\kappa\CZ\NR+\NR\CZ\NG_\kappa)_{l,\,l+j}\,,
\label{expru}
\\
\Nw^{(2)}_j&=&{1\over2}\sum_{l=1}^{N-j}
(\NG_\kappa\NX^0\bar\NV^0-\bar\NV^0\NX^0\NG_\kappa)_{l,\,l+j}\,.
\label{expru0}
\eeqnarray
We first consider $\Nw^{(1)}$.
We note that $\NG_\kappa\CZ\NR+\NR\CZ\NG_\kappa$ is a
bordered c-symmetric matrix and that $\CZ$ is c-symmetric since both $\NZ_1$ and
$\NZ$ are c-symmetric. One thus obtains from (\ref{expru})
\qq
\Nw^{(1)}=\NG_\kappa^{(N-1)}\tilde\CZ,
\label{expru1}
\qqq
where, for $1\leq j\leq N-1$,
\qq
\tilde\CZ_j=\CZ_{1,j+1}.
\label{tildez}
\qqq
In order to compute $\tilde\CZ$,
we note that $\NZ$ solves
the equation $[\NG_\kappa,\NZ]=-{1\over\nu}\{\NR,\NY_1\}$, as is easily checked
from (\ref{matrixZ}) and (\ref{phijY}).
Therefore, $\CZ$ solves, cf.~(\ref{lin3}) and (\ref{decZ}),
\qq
[\NG_\kappa,\CZ]=[\NZ^0,\bar\NV_0]\,,
\label{calz1}
\qqq
and since $\CZ$ is antisymmetric, as both $\NZ_1$ and $\NZ$ are,
it follows from (\ref{matrixZ0}), $(\bar\NV_0)_{ij}=\delta_{ij}g_i$,
and formula (\ref{eforms}), that for $2\leq j\leq N$,
\qq
\CZ_{1j}={1\over2}\sum_{l=1}^{j-1}(g_{j-l}-g_{l})\phi_{j-2l}\,,
\label{solttilz}
\qqq
with the convention $\phi_{-k}=-\phi_{k}$, $0\leq k\leq N-1$.
Thus, $\Nw^{(1)}$ is given by (\ref{expru1}) with
$\tilde\CZ\in{\bf R}^{N-1}$ given by
\qq
\tilde\CZ_{j}={1\over2}\sum_{l=1}^{j}(g_{j+1-l}-g_{l})\phi_{j+1-2l}\,.
\label{ztilde}
\qqq
We next consider $\Nw^{(2)}$. We first note that
\qq
\NG_\kappa\NX^0\bar\NV_0-\bar\NV_0\NX^0\NG_\kappa=
(\NG_{\nu+\kappa}\NX^0\bar\NV_0-\bar\NV_0\NX^0\NG_{\nu+\kappa})+
\nu(\bar\NV_0\NX^0-\NX^0\bar\NV_0),
\label{u2in1}
\qqq
and compute, using (\ref{matrixX0}), (\ref{equphi}), and
$(\bar\NV_0)_{ij}=\delta_{ij}g_i$, that for $i\leq j$,
\qq
(\NG_{\nu+\kappa}\NX^0\bar\NV_0-\bar\NV_0\NX^0\NG_{\nu+\kappa})_{ij}=
\delta_{1i}g_{j}\phi_{j-1}+\delta_{Nj}g_{i}\phi_{N-i}.
\qqq
Therefore,
\qq
(\NG_\kappa\NX^0\bar\NV_0-\bar\NV_0\NX^0\NG_\kappa)_{ij}=
\delta_{i1}g_j\phi_{j-1}+\delta_{jN}g_i\phi_{N-i}
+\nu(g_i-g_j)\phi_{i+j-1}\,,
\label{G3b}
\qqq
with the convention $\phi_{N+k}=-\phi_{N-k}$, $0\leq k\leq N$.
One thus finally obtains for $\Nw^{(2)}\in{\bf R}^{N-1}$,
using in addition that $g_{N-j}=g_{j+1}$,
\qq
\Nw^{(2)}_j=g_{j+1}\phi_{j}+
{\nu\over2}\sum_{l=1}^{N-j}(g_{l}-g_{j+l})\phi_{j-1+2l}\,.
\label{Psi}
\qqq
Using
(\ref{eqphij}), (\ref{solU}), (\ref{expru1}), (\ref{ztilde}), (\ref{Psi}),
and the fact that the
$\phi_j$'s decay exponentially, it is easy to see that $\varphi_1$ is
uniformly bounded in $N$.

\subsection{First-order correction to the temperature profile}

We now analyze the first-order correction to the
temperature profile. It is given by $(\Phi^1_y)_{ii}$ where,
cf.~(\ref{decompPhi}) and (\ref{Philansatz}),
\qq
\Phi_y^1={3k^2T^2\over\omega^4}(\NY_0+\eta\NY_1+\eta^2\NY_2).
\label{decphi4}
\qqq
By (\ref{solzero}), $\NY_0$ does not contribute to $\Phi^1_y$.
In order to compute the diagonal of $\NY_1$, we use
the fact that $\NY_1$ is c-antisymmetric and satisfies the equation,
as a consequence of (\ref{lin2}),
\qq
[\NG_\kappa,\NY_1]=\NG_\kappa\NZ_1\NR+\NR\NZ_1\NG_\kappa+
(\NG_\kappa\NX^0\bar\NV_0-\bar\NV_0\NX^0\NG_\kappa).
\label{G1al}
\qqq
Using (\ref{eformcs}), (\ref{G3b}), and the fact that
$g_{2i}=g_{N-2i+1}$, one thus obtains
for $1\leq i\leq[N/2]$, where $[x]$ denotes the largest integer
smaller or equal to $x$,
\qq
(\NY_{1})_{ii}=
(\NG_\kappa^{(N-1)}\tilde\NZ_{1})_{2i-1}
+\Bigl(g_{2i}\,\phi_{2i-1}+
{\nu\over2}\sum_{l=i}^{N-i}
\phi_{2l}\sum_{k=0}^{i-1}(g_{l-k}-g_{l+k+1})\Bigr),
\label{G10}
\qqq
where $\tilde\NZ_1\in{\bf R}^{N-1}$ is given by
$(\tilde\NZ_{1})_{j}=(\NZ_{1})_{1,j+1}$. Since the $\phi_{j}$ decay exponentially fast
with rate $\alpha$, see (\ref{phi}), it follows that all terms but the first give
an exponentially flat contribution to $(\NY_{1})_{ii}$. We
thus write, and will adopt a similar notation in the sequel,
\qq
(\NY_{1})_{ii}=(\NG_\kappa^{(N-1)}\tilde\NZ_{1})_{2i-1}
+\CO(e^{-\alpha j}).
\label{G1b}
\qqq
In order to compute the dominant term in the above expression, we first use that
$\tilde\NZ_1=\varphi+\tilde\CZ$
where $\tilde\CZ$ is given by (\ref{ztilde}), and
$\NG_{\nu+\kappa}^{(N-1)}\varphi=-\Nw$ where
$\Nw=\NG_\kappa^{(N-1)}\tilde\CZ+\Nw^{(2)}$ with $\Nw^{(2)}$
given by (\ref{Psi}), to obtain
$\tilde\NZ_1=(\NG^{(N-1)}_{\nu+\kappa})^{-1}(\nu\tilde\CZ-\Nw^{(2)})$ and thus
\qq
(\NY_{1})_{ii}=\Bigl((\NG_{\nu+\kappa}^{(N-1)})^{-1}\NG_\kappa^{(N-1)}(\nu\tilde\CZ
-\Nw^{(2)})\Bigr)_{2i-1}
+\CO(e^{-\alpha j}).
\label{tpf1}
\qqq
It follows from the expression (\ref{expru0}) for $\Nw^{(2)}$
and properties of $\NG_\kappa^{(N-1)}$, $\NG_{\nu+\kappa}^{(N-1)}$, and their inverse,
that the second term gives an exponentially flat contribution to the
temperature profile.
To compute the remaining term
$y\equiv\nu(\NG_{\nu+\kappa}^{(N-1)})^{-1}\NG_\kappa^{(N-1)}\tilde\CZ$, we first note that
it satisfies
\qq
\NG_{\nu+\kappa}^{(N-1)}y=\nu\NG_\kappa^{(N-1)}\tilde\CZ.
\label{smallyeq}
\qqq
We next compute $\NG_\kappa^{(N-1)}\tilde\CZ$. In the expression
(\ref{ztilde}) for $\tilde\CZ$, changing the summation index to $k$
with $2k=j+1-2l$ if $j$ is odd and  $2k=j-2l$ if $j$ is even, one
obtains, using in addition the symmetry properties of $g_i$,
that for $j\geq2$
\qq
\tilde\CZ_j=\cases{
\sum_{k=1}^{{j-1\over2}}(g_{{j+1\over2}+k}-g_{{j+1\over2}-k})\phi_{2k}
& if $j$ is odd,\cr
\sum_{k=1}^{{j\over2}}(g_{{j\over2}+k}-g_{{j\over2}+1-k})\phi_{2k-1}
& if $j$ is even.\cr}
\qqq
For $j=1$, $\tilde\CZ_1=0$.
Computing the differences of $g$'s arising in the above expression
leads to
\qq
\tilde\CZ_j={\sinh (N-j)\bar\alpha\over \sinh (N+1)\bar\alpha}
\sum_{k=1}^{{j-1+\bar\jmath}\over2}{\sinh(2k-\bar\jmath)\bar\alpha \over\sinh\bar\alpha}
\phi_{2k-\bar\jmath},
\qqq
where $\bar\jmath=0$ if $j$ is odd and $\bar\jmath=1$ if $j$ is even.
Hence, $\tilde\CZ$ can be rewritten as
\qq
\tilde\CZ_j=\rho_{\bar\jmath}\,{\sinh (N-j)\bar\alpha\over
\sinh(N+1)\bar\alpha}+\CO(e^{-\alpha j}),
\qqq
where the constants $\rho_0$ and $\rho_1$ are given by
\qq
\rho_{\sigma}=\sum_{k=1}^{[N/2]}{\sinh(2k-\sigma)\bar\alpha\over\sinh\bar\alpha}
\phi_{2k-\sigma}\,,\quad
\sigma=0,1.
\qqq
A straightforward computation finally leads to, recalling that
$\cosh\bar\alpha=1+\kappa/2$,
\qq
(\NG_\kappa^{(N-1)}\tilde\CZ)_j=
(-1)^{\bar\jmath+1}\,(2+\kappa)(\rho_1-\rho_0){\sinh (N-j)\bar\alpha\over \sinh
(N+1)\bar\alpha} +C_1\delta_{1j}+\CO(e^{-\alpha j}),
\label{Gztilde}
\qqq
where $C_1$ is a constant that depends on $N$ and $\bar\alpha$ only.
It thus remains to compute the vector $y$ given by equation (\ref{smallyeq}). To this
end, we note that a vector of the form (\ref{Gztilde}) is almost an eigenvector of
$\NG_{\nu+\kappa}^{(N-1)}$. More precisely, one has for $v$ with
$v_j=(-1)^{\bar\jmath+1}\sinh(N-j)\bar\alpha$,
\qq
(\NG_{\nu+\kappa}^{(N-1)} v)_j=(4+\nu+2\kappa)v_j
+\delta_{1j}\,{\sinh N\bar\alpha}.
\label{almostev}
\qqq
Therefore, writing
\qq
y_j=(-1)^{\bar\jmath+1}\,{\nu(2+\kappa)(\rho_1-\rho_0)\over(4+\nu+2\kappa)}
{\sinh (N-j)\bar\alpha\over\sinh (N+1)\bar\alpha}+r_j\,,
\label{smally}
\qqq
and inserting in (\ref{smallyeq}) yield for $r$ the equation
$(\NG_{\nu+\kappa}^{(n-1)}r)_j=C_2\delta_{1j}+\CO(e^{-\alpha j})$ with
$C_2$ a constant depending on $N$ and $\bar\alpha$, cf.~(\ref{Gztilde}) and
(\ref{almostev}), whose solution reads, by using (\ref{equphi}),
\qq
r_j=C_2\phi_j+\CO(e^{-\alpha j}).
\qqq
Hence, $r$ is an exponentially decaying correction to $y$ as given by (\ref{smally}).
Finally, since $(\NY_1)_{ii}=y_{2i-1}$ for
$1\leq i\leq[N/2]$, we obtain from (\ref{smally}),
\qq
(\NY_1)_{ii}=-{\nu(2+\kappa)(\rho_1-\rho_0)\over(4+\nu+2\kappa)}
{\sinh(N+1-2i)\bar\alpha\over\sinh(N+1)\bar\alpha}
+\CO(e^{-2\alpha i}).
\label{diagY1}
\qqq
Since $\NY_1$ is c-antisymmetry, (\ref{diagY1}) also gives the elements $(\NY_1)_{ii}$
for $[N/2]+1\leq i\leq N$.
In particular, since $\cosh\bar\alpha=1+\kappa/2$, it follows that
the contribution of $\NY_1$ to the temperature profile is exponentially
flat in the bulk of the chain whenever $\kappa>0$.
When $\kappa=0$, on the other hand, $\bar\alpha=0$ and
$\NY_1$ gives a linear profile.
In the limit $N\rightarrow\infty$,
it is straightforward to compute that for $\kappa=0$,
$\rho_1$ and $\rho_0$ are given by
\qq
\rho_0={1\over 2\sinh^2\alpha}
\quad{\rm and}\quad
\rho_1={\cosh\alpha\over2\sinh^2\alpha}\,,
\label{rho01}
\qqq
with $\alpha$ defined by $\cosh\alpha=1+\nu/2$.
One thus has $\rho_1-\rho_0=1/(4+\nu)$
and the temperature profile for $\kappa=0$ is given by
\qq
(\NY_1)_{ii}={2\nu\over(4+\nu)^2}\Bigr({2i\over N+1}-1\Bigl)+\CO(e^{-2\alpha i}).
\label{diagYl}
\qqq
The temperature profile is linear, but oriented in the ``wrong'' direction. Indeed, if
for instance
$T_1>T_N$, then one obtains from (\ref{decphi4}),
which involves a multiplication by $\eta=(T_1-T_N)/(T_1+T_N)$,
that the slope is positive.  

\begin{figure}
\begin{center}
\begingroup%
  \makeatletter%
  \newcommand{\GNUPLOTspecial}{%
    \@sanitize\catcode`\%=14\relax\special}%
  \setlength{\unitlength}{0.1bp}%
\begin{picture}(3600,2807)(0,0)%
\includegraphics{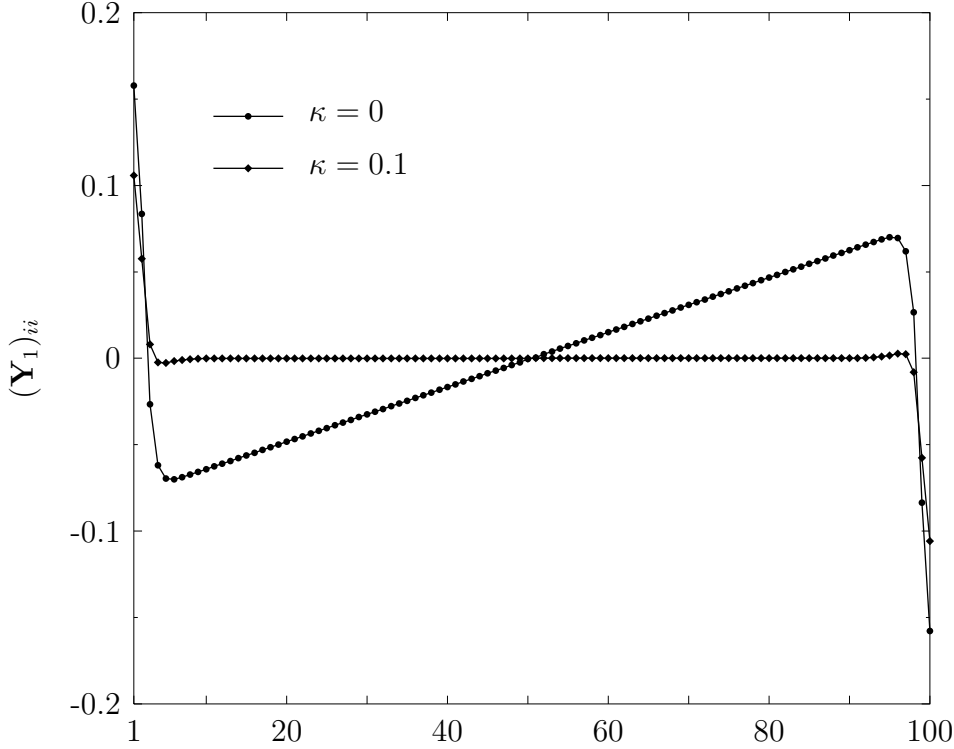}
\put(700,2220){\makebox(0,0)[r]{ }}%
\put(700,2416){\makebox(0,0)[r]{ }}%
\put(1110,2239){\makebox(0,0)[l]{$\kappa=0.1$}}%
\put(1110,2437){\makebox(0,0)[l]{$\kappa=0$}}%
\put(100,1503){%
\makebox(0,0)[b]{\shortstack{$(\NY_1)_{ii}$}}%
}%
\put(3450,100){\makebox(0,0){100}}%
\put(2844,100){\makebox(0,0){80}}%
\put(2238,100){\makebox(0,0){60}}%
\put(1632,100){\makebox(0,0){40}}%
\put(1026,100){\makebox(0,0){20}}%
\put(450,100){\makebox(0,0){1}}%
\put(400,2807){\makebox(0,0)[r]{0.2}}%
\put(400,2155){\makebox(0,0)[r]{0.1}}%
\put(400,1504){\makebox(0,0)[r]{0}}%
\put(400,852){\makebox(0,0)[r]{-0.1}}%
\put(400,200){\makebox(0,0)[r]{-0.2}}%
\end{picture}%
\endgroup
\caption{Contribution of $\NY_1$ to the temperature profile
  ($\nu=1$, $N=100$).}
\end{center}
\end{figure}

We next consider the contribution of
$\NY_2$ to the temperature profile. Since $\NY_2$ is
c-symmetric, it will introduce, if nonzero, a
global shift in the temperature profile. As we shall see, this is
indeed the case.
To compute the diagonal $(\NY_2)_{ii}$, we proceed as for $\NY_1$.
We first recall that $(\NY_2)_{11}=0$, cf. (\ref{Y211zero}), and note that
$\NY_2$ also satisfies,
\qq
[\NG_\kappa,\NY_2]=\NG_\kappa\NZ_2\NR+\NR\NZ_2\NG_\kappa+(\NG_\kappa\NX^0\bar\NV_1-\bar\NV_1\NX^0\NG_\kappa).
\label{GY2}
\qqq
Denoting by $\psi$ the first line of $\NY_2$, i.e.,
\qq
\psi_i\equiv(\NY_2)_{1i}\,,
\qqq
one uses (\ref{eformgen}) to obtain from (\ref{GY2}) the following expression,
for $i\geq2$ and $2i\leq N+1$,
\qq
(\NY_2)_{ii}=\sum_{k=1}^{i-1}\psi_{2k+1}-\sum_{k=1}^{i-1}
\sum_{l=1}^k\NU_{k-l+1,k+l}\,,
\label{NY2a}
\qqq
where $\psi_1=(\NY_2)_{11}=0$ has been used, and
\qq
\NU=\NG_\kappa\NZ_2\NR+\NR\NZ_2\NG_\kappa+(\NG_\kappa\NX^0\bar\NV_1-\bar\NV_1\NX^0\NG_\kappa).
\label{UY2}
\qqq
Since $\NY_2$ is c-symmetric, (\ref{NY2a}) determines all
diagonal elements $(\NY_2)_{ii}$, $2\leq i\leq N-1$.
The first term on the RHS of (\ref{GY2}) is a bordered matrix
and a straightforward computation yields
\qq
\sum_{l=1}^k(\NG_\kappa\NZ_2\NR+\NR\NZ_2\NG_\kappa)_{k-l+1,k+l}=
(\NG_\kappa\zeta)_{2k}\,,
\label{UY21}
\qqq
where $\zeta$ denotes the first line of $\NZ_2$, i.e.,
\qq
\zeta_i=(\NZ_2)_{1i}\,.
\qqq
The second term on the RHS of (\ref{UY2}) is identical to the
corresponding term appearing in (\ref{eqU}), with $\bar\NV_0$ replaced
by the diagonal matrix $(\bar\NV_1)_{ij}=\delta_{ij}\phi_{2i-1}$.
For $1\leq i\leq j\leq N$, it is thus
given by, cf.~(\ref{G3b}),
\beqnarray
(\NG_\kappa\NX^0\bar\NV_1-\bar\NV_1\NX^0\NG_\kappa)_{ij}=
\nu(\phi_{2i-1}-\phi_{2j-1})\phi_{i+j-1}\qquad\qquad\qquad\qquad\nonumber
\\
\qquad\qquad\qquad
+\ \delta_{i1}\phi_{2j-1}\phi_{j-1}
+\delta_{jN}\phi_{2i-1}\phi_{N-i}\,,
\label{UY22}
\eeqnarray
with the convention $\phi_{N+k}=-\phi_{N-k}$, $0\leq k\leq N$.
Inserting (\ref{UY21}) and (\ref{UY22}) into (\ref{NY2a}) leads to
\qq
(\NY_2)_{ii}=\sum_{k=1}^{i-1}\Delta_k
\label{NY2b}
\qqq
where, for $k\geq1$ and $2k\leq N-1$,
\qq
\Delta_k=\psi_{2k+1}-(\NG_\kappa\zeta)_{2k}-
\Bigl(\phi_{2k-1}\phi_{4k-1}+\nu
\phi_{2k}\sum_{l=1}^k\bigl(\phi_{2(k-l)+1}-\phi_{2(k+l)-1}\bigr)\Bigr).
\label{Deltak}
\qqq
One checks that $|\Delta_k|$ decays exponentially.
First, recalling (\ref{phi}) and our convention
$\phi_{N+k}=-\phi_{N-k}$, $0\leq k\leq N$, this is clearly true of the
last two terms in (\ref{Deltak}).
Next, an expression for the first line of $\NY_2$ can be obtained from
equation (\ref{qua3}) by using that $\NZ_2$ is c-antisymmetric.
Formula (\ref{eformcs})
and $(\NZ_2)_{k,k+1}=0$, cf.~(\ref{cur2}), imply that for
$1\leq k\leq[(N-1)/2]$,
\qq
{1\over\nu}
\psi_{2k+1}={1\over2}\sum_{n=1}^k\phi_{2n}
\sum_{l=k}^{N-k-1}(\phi_{2(l+n)+1}-\phi_{2(l-n)+1})\,,
\label{psiodd}
\qqq
with the convention $\phi_{N+k}=-\phi_{N-k}$, $0\leq k\leq N$.
In particular, $\psi_{2k+1}$ decays 
exponentially. We finally compute $\zeta$, the first line of 
$\NZ_2$. One has $\zeta_1=\zeta_N=0$ by antisymmetry and c-antisymmetry of $\NZ_2$,
and applying formula (\ref{eformscs}) to equation (\ref{qua3})
yields for $2\leq j\leq N-1$
\qq
\zeta_j={1\over4}
\sum_{n=1}^{j-1}\phi_{j-2n}\sum_{l=1}^{N-j}(\phi_{2(l+n)-1}-\phi_{2(j+l-n)-1}),
\qqq
with the conventions $\phi_{-k}=-\phi_k$
and $\phi_{N+k}=-\phi_{N-k}$, $0\leq k\leq N$.
Therefore, one has for $2\leq i\leq[(N+1)/2]$,
\qq
(\NY_2)_{ii}=h+\CO(e^{-\alpha i}),
\label{NY2c}
\qqq
where the constant $h$ is given by
\qq
h=h_1+\nu h_2,
\label{Ch}
\qqq
with
\beqnarray
h_1&=&\sum_{k=1}^{[{N-1\over2}]}\Bigl(2\zeta_{2k+1}-(2+\kappa)\zeta_{2k}
-\phi_{2k-1}\phi_{4k-1}\Bigr),
\label{Ch1}
\\
h_2&=&\sum_{k=1}^{[{N-1\over2}]}\Bigl({1\over\nu}\psi_{2k+1}
-\phi_{2k}\sum_{l=1}^k\bigl(\phi_{2(k-l)+1}-\phi_{2(k+l)-1}\bigr)\Bigr).
\label{Ch2}
\eeqnarray
A straightforward, but lengthy, computation yields the
following asymptotic formulas for large $N$,
\beqnarray
h_1&=&{\cosh\alpha(\cosh\alpha-1-\kappa/2)\over2e^{\alpha}\sinh^2\alpha\sinh3\alpha}\,,
\label{Ch1as}
\\
h_2&=&-{1\over4\sinh^2\alpha}
\Bigl({1\over\cosh\alpha}+{\cosh\alpha\over e^\alpha\sinh3\alpha}\Bigr)\,.
\label{Ch2as}
\eeqnarray
Recalling that $\cosh\alpha=1+(\nu+\kappa)/2$, one obtains 
\qq
h=-{2\nu\over(\nu+\kappa)(2+\nu+\kappa)(4+\nu+\kappa)}\,.
\label{hasympt}
\qqq

\begin{figure}
\begin{center}
\begingroup%
  \makeatletter%
  \newcommand{\GNUPLOTspecial}{%
    \@sanitize\catcode`\%=14\relax\special}%
  \setlength{\unitlength}{0.1bp}%
\begin{picture}(3600,2807)(0,0)%
\includegraphics{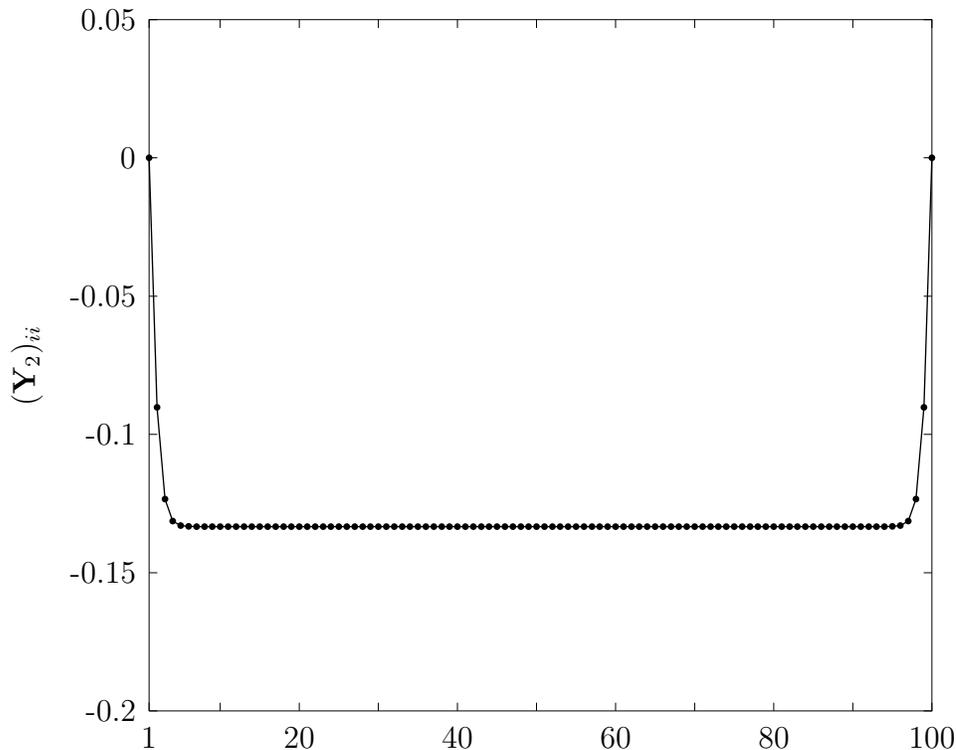}
\put(100,1503){%
\makebox(0,0)[b]{\shortstack{$(\NY_2)_{ii}$}}%
}%
\put(3450,100){\makebox(0,0){100}}%
\put(2854,100){\makebox(0,0){80}}%
\put(2258,100){\makebox(0,0){60}}%
\put(1662,100){\makebox(0,0){40}}%
\put(1066,100){\makebox(0,0){20}}%
\put(500,100){\makebox(0,0){1}}%
\put(450,2807){\makebox(0,0)[r]{0.05}}%
\put(450,2286){\makebox(0,0)[r]{0}}%
\put(450,1764){\makebox(0,0)[r]{-0.05}}%
\put(450,1243){\makebox(0,0)[r]{-0.1}}%
\put(450,721){\makebox(0,0)[r]{-0.15}}%
\put(450,200){\makebox(0,0)[r]{-0.2}}%
\end{picture}%
\endgroup
\caption{Contribution of $\NY_2$ to the temperature profile
  ($\nu=1$, $N=100$).}
\end{center}
\end{figure}

\vskip 0.3cm
{\bf Acknowledgments}
\vskip 0.3cm
\noindent
We thank K.~Aoki, A.~Kupiainen, L.~Rey-Bellet, H.~Spohn, H.~Tasaki,
N.~Yoshida and E.~Zabey for useful discussions
during the preparation of this work.
R.L.~thanks T.~Shiota for his hospitality at Kyoto University.

\end{document}